\documentclass[11pt,a4paper]{article}

\usepackage[truedimen,margin=34mm]{geometry} 

\usepackage{mathrsfs}
\usepackage{amssymb}
\usepackage{amsmath}
\usepackage{ascmac}
\usepackage{amsthm}
\usepackage[pdftex]{graphicx}
\usepackage[pdftex]{color}
\usepackage{natbib}
\usepackage{color}
\usepackage{setspace}
\usepackage{url}

\usepackage{titlesec}
\titleformat*{\section}{\large\bfseries}
\titleformat*{\subsection}{\it}

\newtheorem{algo}{Algorithm}

\def\ep{{\varepsilon}}

\def\th{{\theta}}

\def\thh{{\widehat{\th}}}
\def\gh{{\widehat{g}}}

\def\pih{{\widehat{\pi}}}

\newcommand{\argmax}{\mathop{\rm arg~max}\limits}

\title{{\bf Spatially Clustered Regression}}
\date{}

\begin{document}

\maketitle
\doublespacing

\vspace{-1.5cm}
\begin{center}
{\large Shonosuke Sugasawa$^{*}$\footnote{Corresponding author, Address: 5-1-5, Kashiwanoha, Kashiwa, Chiba 2778568, JAPAN, Email: sugasawa@csis.u-tokyo.ac.jp} and Daisuke Murakami$^{\dagger}$}
\end{center}

\medskip
\noindent
$^{*}$Center for Spatial Information Science, The University of Tokyo\\
$^{\dagger}$Department of Data Science, The Institute of Statistical Mathematics

\medskip
\medskip
\begin{center}
{\bf \large Abstract}
\end{center}

\vspace{-0cm}
Spatial regression or geographically weighted regression models have been widely adopted to capture the effects of auxiliary information on a response variable of interest over a region. In contrast, relationships between response and auxiliary variables are expected to exhibit complex spatial patterns in many applications. This paper proposes a new approach for spatial regression, called spatially clustered regression, to estimate possibly clustered spatial patterns of the relationships.
We combine K-means-based clustering formulation and penalty function motivated from a spatial process known as Potts model for encouraging similar clustering in neighboring locations.
We provide a simple iterative algorithm to fit the proposed method, scalable for large spatial datasets.  
Through simulation studies, the proposed method demonstrates its superior performance to existing methods even under the true structure does not admit spatial clustering. 
Finally, the proposed method is applied to crime event data in Tokyo and produces interpretable results for spatial patterns.
The R code is available at \url{https://github.com/sshonosuke/SCR}.


\newpage
\section{Introduction}
Spatial heterogeneity, which is often referred to as the Second Law of Geography \citep{goodchild2004validity}, is ubiquitous in spatial science. Geographically weighted regression \citep[GWR;][]{GWRpaper, GWRbook}, which is a representative approach for modeling spatial heterogeneity, has widely been adopted for modeling possibly spatially varying regression coefficients; its applications cover social science \cite[e.g.][]{hu2016spatially}, epidemiology \citep[e.g.][]{nakaya2005geographically} and environmental science \cite[e.g.][]{zhou2019application}.

Despite the success, GWR is known to be numerically unstable and may produce extreme estimates of coefficients \citep[e.g.][]{wheeler2005multicollinearity,cho2009extreme}. To address the drawback, a wide variety of regularized GWR approaches have been developed \citep[e.g.][]{wheeler2007diagnostic, wheeler2009simultaneous, barcena2014alleviating}. More recently, \cite{comber2016local} considered local regularization to enhance accuracy and stability. Still, it is less clear how to regularize GWR to improve stability while maintaining its computational efficiency. Bayesian spatially varying coefficient model \citep{Gelfand2003,Finley2011} is another popular approach for modeling spatial heterogeneity in regression coefficients. 
While \cite{wheeler2009comparing} and \cite{wolf2018single} among others have suggested its stability and estimation accuracy, this approach can be computationally very intensive for large samples, limiting applications of spatial regression techniques to modern large spatial datasets.  
Therefore, an alternative method that has stable estimation performance, as well as computational efficiency under large datasets, is strongly required.

This paper proposes a new effective approach for spatial regression with possibly spatially varying coefficients or non-stationarity.
Our fundamental idea is a combination of regression modeling and clustering; we assume all the geographical locations can be divided into a finite number of groups, where locations in the same groups share the same regression coefficients. Hence, possibly smoothed surfaces of varying regression coefficients are approximated by step functions.
Owing to the clustering technique, the estimation results would be numerically stable and more accessible to interpret than GWR.
The idea to incorporate spatial clustering into regression is not new.
There have been some two-stage procedures \citep[e.g.][]{anselin1990spatial,bille2017two,lee2017cluster,nicholson2019spatial}, but they tend to be ad-hoc combinations of clustering and regression.
In contrast, the proposed method carries out regression and clustering simultaneously, which can produce reasonable spatial clustering depending on regression structures.

To introduce such a spatial clustering nature, we employ indicators showing the group to which the corresponding location belongs, and we estimate the grouping parameters and group-wise regression models simultaneously.
For estimating group memberships, it would be reasonable to impose that the geographically neighboring locations are likely to belong to the same groups. To this end, we introduce a penalty function to encourage such spatially clustered structures motivated from the hidden Potts model \citep{Potts1952} that was originally developed for modeling spatially correlated integers. 
We will demonstrate that the proposed objective function can be easily optimized by a simple iterative algorithm similar to $K$-means clustering.
In particular, updating steps in each iteration do not require computationally intensive manipulations, so that the proposed algorithm is much more scalable than GWR.
For selecting the number of groups $G$, we employ an information criterion. 
Moreover, the proposed approach allows substantial extensions to include variable selection or semiparametric additive modeling, which cannot be achieved by existing techniques such as GWR.

Recently, sophisticated statistical methods combining regression modeling and clustering have been studied in the literature.
In the context of spatial regression, \cite{Li2019} and \cite{zhao2020solution} adopted a fused lasso approach to shrink regression coefficients in neighboring areas toward $0$, which results in spatially clustered regression coefficients.
However, the computation cost under large datasets is substantial, and the performance is not necessarily reasonable, possibly because the method does not take account of spatially heterogeneous variances, which will be demonstrated in our numerical studies.    
On the other hand, in the context of panel data analysis, clustering approaches using grouping indicators like the proposed method have been widely studied \citep[e.g.][]{Bon2015, Wang2018, Ito2020}. Still, the existing works did not take account of spatial similarities among the grouping indicators.

This paper is organized as follows.
In Section \ref{sec:method}, we introduce the proposed methods, estimation algorithms and discuss some related issues.  
In Section \ref{sec:sim}, we evaluate the numerical performance of the proposed methods together with some existing methods through simulation studies.
In Section \ref{sec:app}, we demonstrate the proposed method through spatial regression modeling of the number of crimes in the Tokyo metropolitan area.
Finally, we give some discussions in Section \ref{sec:disc}.

\section{Spatially Clustered Regression}\label{sec:method}

\subsection{Models and estimation algorithm}
Let $y_i$ be a response variable and $x_i$ is a vector of covariates in the $i$th location, for $i=1,\ldots,n$, where $n$ is the number of samples.
We suppose we are interested in the conditional distribution $f(y_i|x_i; \theta_i,\psi)$, where $\theta_i$ and $\psi$ are vectors of unknown parameters.
Here $\theta_i$ may change over different locations and represent spatial heterogeneity while $\psi$ is assumed constant in all the areas.
For example, $f(y_i|x_i; \theta_i,\psi)=\phi(y_i;x_{i1}^t\theta_i+x_{i2}^t\gamma,\sigma^2)$ with $x_i=(x_{i1},x_{i2})$ and $\psi=(\gamma,\sigma^2)$.
We assume that location information $s_i$ (e.g. longitude and latitude) is also available for the $i$th location.
In what follows, we assume that there is no static parameter $\psi$ for simplicity, and all the results given above can be easily extended to the case.

Without any structures for $\theta_i$, we cannot identify these parameters since a repeated measurement on the same location is rarely available in practice.
Hence, we assume that $n$ locations are divided into $G$ groups, and locations in the same group share the same parameter values of $\theta_i$.
For a while, we treat $G$ as a fixed value, but a data-dependent selection of $G$ will be discussed later.
We introduce $g_i\in \{1,\ldots,G\}$, an unknown group membership variable for the $i$th location, and let $\theta_i=\theta_{g_i}$.
Then, the distinct values of $\theta_i$'s reduce to $\theta_1,\ldots,\theta_G$, where $\theta=(\theta_1^t,\ldots,\theta_G^t)^t$ is the set of unknown parameters.
Therefore, the unknown parameters in the model is the structural parameter $\theta$ and the membership parameter $g=(g_1,\ldots,g_n)^t$.

Regarding the membership parameter, it would be reasonable to consider that the membership in neighboring locations is likely to have the same memberships, which means that the fitted conditional distributions are likely to be the same in the adjacent locations.
To encourage such a structure, we introduce a penalty function motivated by a spatial process for discrete space known as the Potts model \citep{Potts1952}.
The same penalty function is first adopted in \cite{Sugasawa2020} in mixture modeling. 
The joint probability function of the Potts model is given by
$$
\pi(g_1,\ldots,g_n|\phi)\propto \exp\left(\phi\sum_{i<j}w_{ij}I(g_i=g_j)\right),
$$
where $w_{ij}=w(s_i,s_j)\in[0,1]$, $w(\cdot,\cdot)$ is a weighting function, and $\phi$ controls strength of spatial correlation.
Note that the normalizing constant in the above distribution is not tractable. Still, we treat $\phi$ as a fixed tuning parameter rather than an unknown parameter so that we do not have to deal with the normalizing constant in the following argument.   
Since the conditional distribution of $g_i$ given other variables is the same form as one given above, the conditional distribution put more weights on $w_{ij}I(g_i=g_j)$ as $\phi$ is larger.
Then, we propose the following penalized likelihood:
\begin{equation}\label{PL}
Q(\theta, g)\equiv \sum_{i=1}^n\log f(y_i|x_i;\theta_{g_i})+\phi\sum_{i<j}w_{ij}I(g_i=g_j).
\end{equation}
The above objective function can be regarded as the logarithm of the joint distribution function of $y_1,\ldots,y_n$, and $g$.
We define the estimator of $\theta$ and $g$ as the maximizer of the objective function $Q(\theta, g)$.

For maximizing the objective function (\ref{PL}), we can employ a simple iterative algorithm similar to $K$-means clustering, which iteratively updates the membership variables $g$ and the other parameters. Each updating step is straightforward since the maximization of the objective function (\ref{PL}) given $g$ is the same as maximizing the log-likelihood function based on samples classified to each group.
The detailed algorithm is given as follows:

\medskip
\noindent
\begin{algo} \ (Spatially clustered regression)
\label{algo:SCR}
\begin{itemize}
\item[1.]
Set initial values $\theta_{(0)}$ and $g_{(0)}$.

\item[2.]
Update the current parameter values $\theta_{(k)}$ and $g_{(k)}$ as follows:

\begin{itemize}
\item
Update the group-wise parameter $\theta_g$ separately for $g=1,\ldots,G$: 
$$
\theta^{(k+1)}_g=\argmax_{\theta_g} \sum_{i=1}^nI(g_i^{(k)}=g)\log f(y_i|x_i;\theta_g).
$$

\item
Update the membership variable: 
$$
g_i^{(k+1)}=\argmax_{g\in \{1,\ldots,G\}}\bigg\{\log f(y_i|x_i;\theta^{(k+1)}_g)+\phi\sum_{j=1; j\neq i}^nw_{ij}I(g=g_j^{(k)})\bigg\}.
$$

\end{itemize}

\item[3.]
Repeat the step 2 until convergence.
\end{itemize}
\end{algo}

\medskip
Note that the updating step for $\theta_g$ is easy when $f(y_i|x_i;\theta_g)$ is a standard  regression model .
For example, when $f(y_i|x_i;\theta_g)$ is a Gaussian linear regression model, the updating process for $\theta_g$ are obtained in closed forms.
On the other hand, for updating $g_i$, we just need to calculate values of the penalized likelihood function for all $g\in \{1,\ldots, G\}$, separately for each $i$, which is not computationally intensive as long as $G$ is moderate.
Therefore, each updating step is relatively easy to carry out and computationally less intensive. 
The convergence in the algorithm is monitored by the difference between the current values and updated values. The algorithm should be terminated when the difference is smaller than the user-specified tolerance value $\ep$, where we used $\ep=10^{-6}$ in our numerical studies.

\subsection{Fuzzy clustered regression}
Although Algorithm \ref{algo:SCR} produce interpretable results due to its clustering property, the spatially clustered structure could be restrictive in terms of estimation accuracy when region-wise constant functions cannot reasonably approximate the underlying structure.
To overcome the difficulty, we consider the smoothed version of the proposed method by incorporating fuzzy clustering, which allows the uncertainty of clustering by introducing smoothed weight determined by the likelihood function.
Specifically, we consider the following synthetic probability that the $i$th location belongs to group $g$ given the other group membership variables:
$$
\pi_{ig}=\frac{\big[f(y_i|x_i;\theta_g)\exp\{\phi\sum_{j=1; j\neq i}^nw_{ij}I(g=g_j)\}\big]^{\delta}}{\sum_{g'=1}^G \big[f(y_i|x_i;\theta_{g'})\exp\{\phi\sum_{j=1; j\neq i}^nw_{ij}I(g'=g_j)\}\big]^{\delta}},
$$ 
where $\delta$ controls the degree of fuzziness.
As $\delta\to\infty$, the maximum probability among $\{\pi_{i1},\ldots,\pi_{iG}\}$ converges to $1$, resulting in the same hard clustering as in SCR. 
We use $\delta=1$ as a default choice since $\pi_{ig}$ can be seen as the conditional probability of $g_i=g$ given the data. 
The iterative algorithm is given as follows.

\medskip
\noindent
\begin{algo} \ (Spatially fuzzy clustered regression)
\label{algo:SFCR}
\begin{itemize}
\item[1.]
Set initial values $\theta^{(0)}$ and $g^{(0)}$.

\item[2.]
Compute the following weights for $i=1,\ldots,n$ and $g=1,\ldots,G$.
$$
\pi_{ig}^{(k)}=
\frac{\left[f(y_i|x_i;\theta_g^{(k)})\exp\big\{\phi\sum_{j=1; j\neq i}^nw_{ij}I(g=g_j^{(k)})\big\}\right]^{\delta}}
{\sum_{g'=1}^G \left[f(y_i|x_i;\theta_{g'}^{(k)})\exp\big\{\phi\sum_{j=1; j\neq i}^nw_{ij}I(g'=g_j^{(k)})\big\}\right]^{\delta}}.
$$

\item[3.]
Update the current parameter values $\theta_{(k)}$ and $g_{(k)}$ as follows:

\begin{itemize}

\item
Update the group-wise parameter $\theta_g$ separately for $g=1,\ldots,G$: 
$$
\theta^{(k+1)}_g=\argmax_{\theta_g} \sum_{i=1}^n\pi_{ig}^{(k)}\log f(y_i|x_i;\theta_g).
$$

\item
Update the membership variable: $g_i^{(k+1)}={\rm argmax}_{g\in \{1,\ldots,G\}}\pi_{ig}^{(k)}$.

\end{itemize}

\item[4.]
Repeat the steps 2 and 3 until convergence.
\end{itemize}
\end{algo}

\medskip
Note that the updating step for $\theta_g$ corresponds to maximizing the weighted objective function, which is easy for typical regression models.
The updating processes for $\pi_{ig}$ and $g_i$ can also be easily carried out without any computational difficulty.
Based on the outputs from Algorithm \ref{algo:SFCR}, we can compute the smoothed estimator of $\th_i$ as $\thh_i=\sum_{g=1}^G \pih_{ik}\thh_g$.
Although this smoothed estimator does not hold a clustering nature due to area-wise mixing rates $\pih_{ik}$, it can flexibly adapt to local changes of underlying spatially varying parameters.

\subsection{Selection of tuning parameters}

In the proposed method, we have two tuning parameters, $G$, the number of groups, and $\phi$ controlling the strength of spatial dependence of $ g_i$s.
Since we found that the specific choice of $\phi$ is not very sensitive as long as $\phi$ is strictly positive, thereby we simply recommend setting $\phi=1$.
Although the number of groups, $G$, could be determined according to the prior information regarding the dataset, a data-dependent method can be employed by using the following information criterion: 
\begin{equation}\label{IC}
{\rm IC}(G)=-2\sum_{i=1}^n \log f(y_i|x_i;\widehat{\th}_{\hat{g}_i})+c_n {\rm dim}(\th),
\end{equation}
where $c_n$ is a constant depending on the sample size $n$ and ${\rm dim}(\th)$ denotes the dimension of $\theta$ which depends on $G$. 
Specifically, we use $c_n=\log n$, which leads to a BIC-type criterion. 
We select a suitable value of $G$ as $\widehat{G}={\rm argmin}_{G\in \{G_1,\ldots,G_L\}}{\rm IC}(G)$, where $G_1,\ldots, G_L$ are candidates of $G$.

\subsection{Estimation in locations without samples}
Suppose we want to estimate $\theta_r$ at some location $r$ without samples.
Since there is no data point at location $r$, there is no likelihood based on the data, and the grouping parameter $g_r$ under SCR can be simply estimated as  
$$
\gh_r=\argmax_{g\in \{1,\ldots,G\}}\sum_{i=1}^nw_{ri}I(g=\gh_i),
$$
which results in $\thh_r=\thh_{\gh_r}$.
In a similar way, estimation in location $r$ without samples under SFCR can be performed as $\thh_r=\sum_{g=1}^G \pih_{rg}\thh_g$, where 
$$
\pih_{rg}=
\frac{\left[\exp\big\{\phi\sum_{i=1}^nw_{ij}I(g=\gh_i)\big\}\right]^{\delta}}
{\sum_{g'=1}^G \left[\exp\big\{\phi\sum_{i=1}^nw_{ij}I(g'=\gh_i)\big\}\right]^{\delta}}.
$$

\subsection{Computation of standard errors}
For evaluating the uncertainty of the final estimator $\thh_i$, we here propose two approaches. 
The first approach is a somewhat crude method that computes the standard errors of $\thh_g$ based on the model $f(y|x;\theta_g)$ using samples with $\gh_i=g$, which is easily performed as long as the model is tractable. 
However, this procedure ignores the estimation error in $\gh_i$; thereby, the calculated standard errors may underestimate the true ones. 
The second procedure is a computationally demanding but valid procedure using the parametric bootstrap.
We first generate bootstrap samples $y_i^{\ast}$ from the estimated model, $f(\cdot|x_i;\thh_{\gh_i})$, and apply SCR or SFCR to the bootstrap samples to get the bootstrap estimators, $\thh_g^{\ast}$ and $\gh_i^{\ast}$.
Since the bootstrap procedure requires fitting SCR or SFCR to each replication of the bootstrap samples, it can be computationally intensive under large spatial data. 
However, the parametric bootstrap would be feasible in practice due to the efficient and scalable optimization algorithm.

\section{Simulation Studies}\label{sec:sim}

\subsection{Simulation settings}

We present simulation studies to illustrate the performance of the proposed spatially clustered regression (SCR) and spatially fuzzy clustered regression (SFCR) methods under two scenarios for underlying structures of regression coefficients. 
In both scenarios, we uniformly generated $n=1000$ spatial locations $s_1,\ldots,s_n$ in the domain $\{s=(s_1,s_2)\ | \ s_1\in [-1,1], \ s_2\in [0,2], \ s_1^2+0.5s_2^2>(0.5)^2\}$.
Then, we generated two covariates from spatial processes, following \cite{Li2019}.
Let $z_1(s_i)$ and $z_2(s_i)$ be the two independent realizations of a spatial Gaussian process with mean zero and a covariance matrix defined from an isotropic exponential function: ${\rm Cov}(z_k(s_i), z_k(s_j))=\exp(-\|s_i-s_j\|/\eta)$, $k=1,2$, where $\eta$ is the range parameter. 
We considered three cases of the parameter, $\eta=0.2, 0.6, 1$, which are referred to as weak, moderate, and strong spatial correlation.  
Then, we define two covariates $x_1(s_i)$ and $x_2(s_i)$ via linear transformations $x_1(s_i)=z_1(s_i)$ and $x_2(s_i)=rz_1(s_i)+\sqrt{1-r^2}z_2(s_i)$ with $r=0.75$, which allows dependence between $x_1(s_i)$ and $x_2(s_i)$. 
Then, the response at each location is generated from the following model: 
$$
y(s_i) = \beta_0(s_i) + \beta_1(s_i)x_1(s_i) + \beta_2(s_i)x_2(s_i) +\sigma(s_i)\ep(s_i), \ \ \ i=1,\ldots,n,
$$
where $\ep(s_i)$'s are mutually independent and $\ep(s_i)\sim N(0,1)$.
Regarding the settings of the regression coefficients and error variance, we considered the following two scenarios:

\begin{itemize}
\item[-]
{\bf (Scenario 1: Spatially clustered parameters)} \ \
The sampled domain is divided into 6 regions $D_{jk}=\{s \ | \ g_{1j}<s_1\leq g_{1,j+1}, \ g_{2k}<s_2\leq g_{2,k+1}\}$ for $j=0, 1$ and $k=0, 1, 2$, where $g_{1j}=-1+j$ and $g_{2k}=2k/3$. 
Regression coefficients and error variance for locations in $D_{jk}$ are set as follows: 
\begin{align*}
&\beta_0(s_i)=2(g_{1j}+g_{2k}), \ \ \ \ 
\beta_1(s_i)=g_{1j}^2+g_{2k}^2, \\
&\beta_2(s_i)=-g_{1j}-g_{2k}, \ \ \ \ 
\sigma(s_i)=0.5+0.2|g_{1j}-g_{2k}|,
\end{align*}  
thereby the regression coefficients and error variance are constant within the region $D_{jk}$.

\item[-]
{\bf (Scenario 2: Spatially smoothed parameters)}\ \ 
Each regression coefficient was independently generated from a Gaussian spatial process. 
We set that all the processes have a zero mean and isotropic exponential function given by 
$$
{\rm Cov}(\beta_k(s_i), \beta_k(s_j))=\tau^2\exp\left(-\frac{\|s_i-s_j\|}{\psi_k}\right), \ \ \ \ k=0, 1, 2, 
$$
where $\psi_k$ is the range parameter and $\tau^2$ is the variance parameter. 
We fix $\tau^2=2$ and $\psi_k=k+1$ in our study. 
Regarding the error variance, we set $\sigma(s_i)=0.2\exp(u(s_i))$, where $u(s_i)$ is a zero mean Gaussian spatial process with the same isotropic exponential function ${\rm Cov}(u_k(s_i), u_k(s_j))=(0.5)^2\exp(-\|s_i-s_j\|/3)$.

\end{itemize}

\subsection{Methods}\label{sec:sim-method}
For the simulated dataset, we applied the proposed SCR with $\phi=1$ and the number of groups $G$ selected among $\{5, 10, \ldots, 30\}$ by using the BIC-type criterion (\ref{IC}). 
We also applied SFCR with $\delta=1$ and the same selected value of $G$ in SCR. 
Regarding the weight $w_{ij}$, we adopted two cases; five nearest neighbor (for the $i$th location we set $w_{ij}=1$ for the five nearest locations and $w_{ij}=0$ otherwise) and exponential weight function, namely, $w_{ij}=\exp(-\|s_i-s_j\|^2/(0.1)^2)$, which are denoted by -n and -e, respectively. 
For competitors, we adopted two methods.
The first one is geographically weighted regression (GWR) as the most standard method in spatial regression. 
Although spatially varying coefficient models \citep{Gelfand2003} are also standard methods, previous studies suggest that spatially varying coefficient models tend to produce similar results to those of GWR \citep{Finley2011} since they can be regarded as a model-based version of GWR. 
Therefore, we only adopted GWR in this study.
The bandwidth parameter in GWR was chosen via cross-validation, and all the estimation procedure was carried out via R package ``spgwr" \citep{spgwr}, in which Gaussian kernel is used as the spatial weight function.
We also applied the multiscale GWR \citep{fotheringham2017multiscale}, which estimates the bandwidth parameter for each covariate, as an advanced version of GWR by using R package ``GWmodel" \citep{gollini2015gwmodel}.
The second competitor is a more advanced and recent regularization technique called spatial homogeneity pursuit (SHP) proposed in \cite{Li2019}.
In this method, we first constructed a minimum spanning tree connecting all the locations using R package ``ape" \citep{ape}, and then lasso regularized estimation is applied using the R package "glmnet" \citep{glmnet}. 
Following \cite{Li2019}, the tuning parameter in the regularized estimation was selected by the BIC-type criterion.

The estimation performance is evaluated based on the mean squared error (MSE) given by 
$$
{\rm MSE}=\frac1{np}\sum_{i=1}^n\sum_{k=0}^{p-1}\left\{\widehat{\beta}_k(s_i)-\beta_k(s_i)\right\}^2,
$$ 
where $p=3$ and $\widehat{\beta}_k(s_i)$ is the estimated value of $\beta_k(s_i)$. 
We also evaluate the performance in terms of spatial interpolation (estimation in locations without samples) by generating $m=100$ additional locations and true regression coefficients. 
Using GWR, SCR, and SFCR, we obtained estimates in the locations without samples and assessed the performance via the same MSE except for $n$ replaced with $m$.

\subsection{Results}
\begin{figure}[!htb]
\centering
\includegraphics[width=12cm,clip]{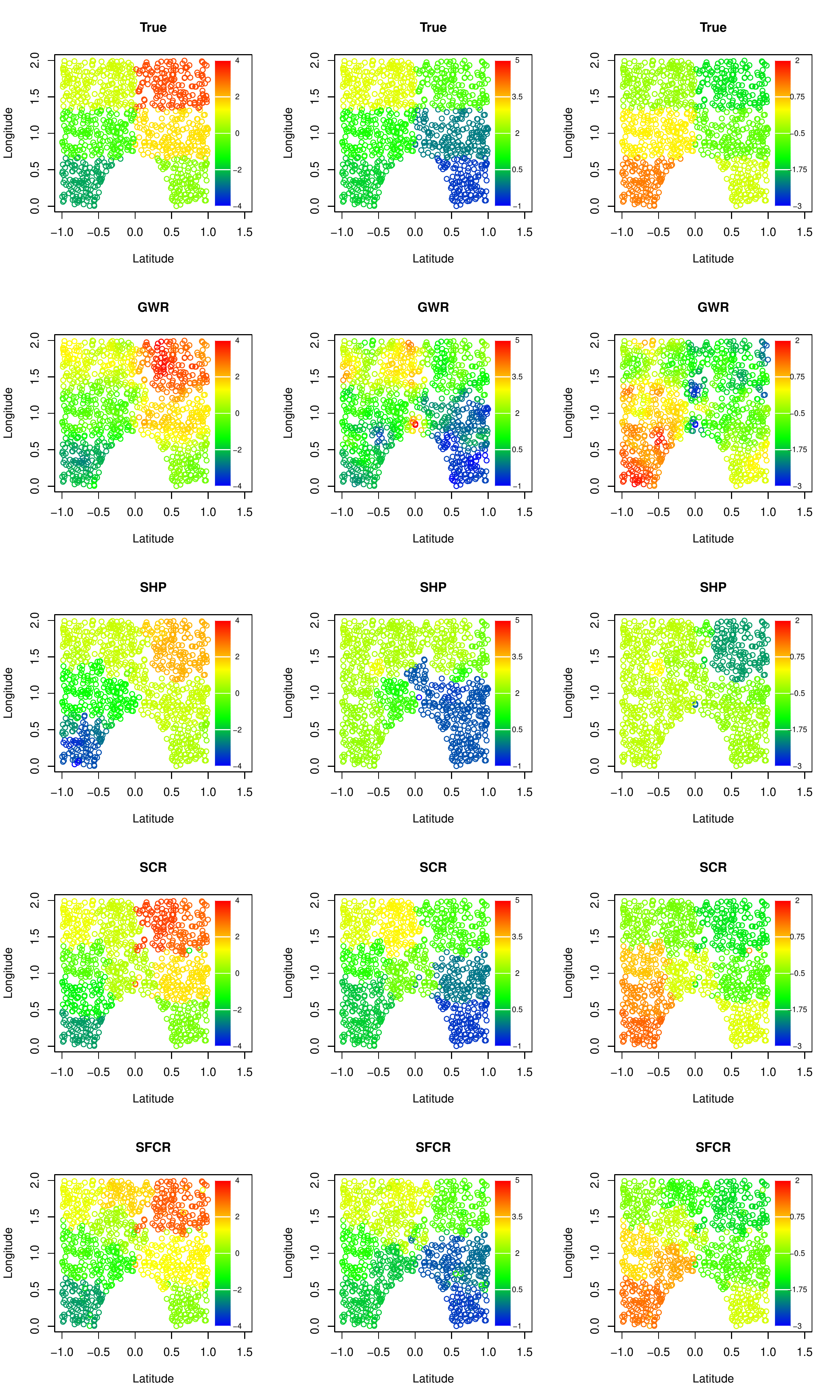}
\caption{Scenario 1: spatial patterns of true and estimated regression coefficients based on GWR, SHP and SCR in one simulation with the spatial range parameter $\eta=0.2$ for covariates.
The left, center and right columns correspond to $\beta_0$, $\beta_1$ and $\beta_2$, respectively. 
\label{fig:sim1}
}
\end{figure}

\begin{figure}[!htb]
\centering
\includegraphics[width=12cm,clip]{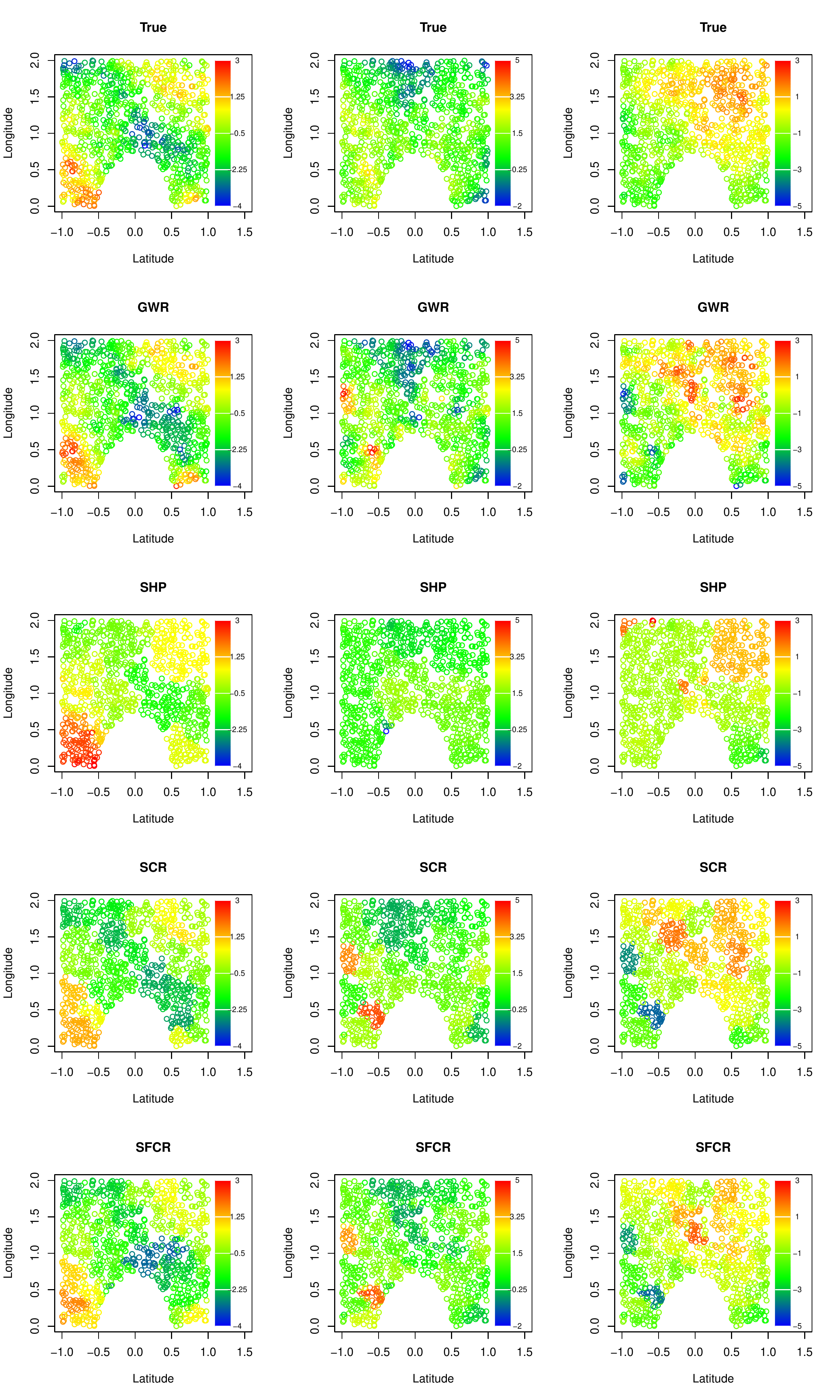}
\caption{Scenario 2: spatial patterns of true and estimated regression coefficients based on GWR, SHP, SCR, SFCR in one simulation with the spatial range parameter $\eta=0.2$ for covariates.
The left, center and right columns correspond to $\beta_0$, $\beta_1$ and $\beta_2$, respectively. 
\label{fig:sim2}
}
\end{figure}

We first show the result using a single simulated dataset with $\eta=0.2$. 
In Table \ref{tab:CPT}, we reported the computation time (second) of each method, where the program was run on a PC with a 3 GHz 8-Core Intel Xeon E5 8 Core Processor with approximately 16GB RAM.
It is observed that the proposed method is computationally comparable with GWR, whereas SHP is computationally much more intensive than the other methods. 
In the SHP method, we found that computation time for the minimum spanning tree accounts for a large portion of the total computation time of SHP.    
The spatial patterns of the true and estimated parameter values are presented in Figures \ref{fig:sim1} and \ref{fig:sim2}. 
In scenario 1, the proposed method can successfully capture the underlying clustered structures of the regression coefficients and detects the abrupt changes across the boundaries of adjacent clusters.
Note that the selected number of groups was $G=10$, which is the smallest choice among candidates that are larger than the true number of clusters.
On the other hand, GWR does not provide estimates having clustered structures and produces poor estimates in some locations. 
Although SHP can capture similar clustered structures, the proposed SCR can more precisely capture the structure.  
In scenario 2, it is observed that GWR can precisely estimate the spatially smoothed regression coefficients while SCR can also produce reasonable estimates by allowing a large number of groups. 
In fact, the largest number ($G=30$) among the candidates was selected in this scenario.  
Moreover, SFCR produces more smoothed estimates than SCR, which are more similar to ones by GWR. 
In contrast, the results of SHP are not necessarily satisfactory compared with the other methods.

\begin{figure}[!htb]
\centering
\includegraphics[width=14cm,clip]{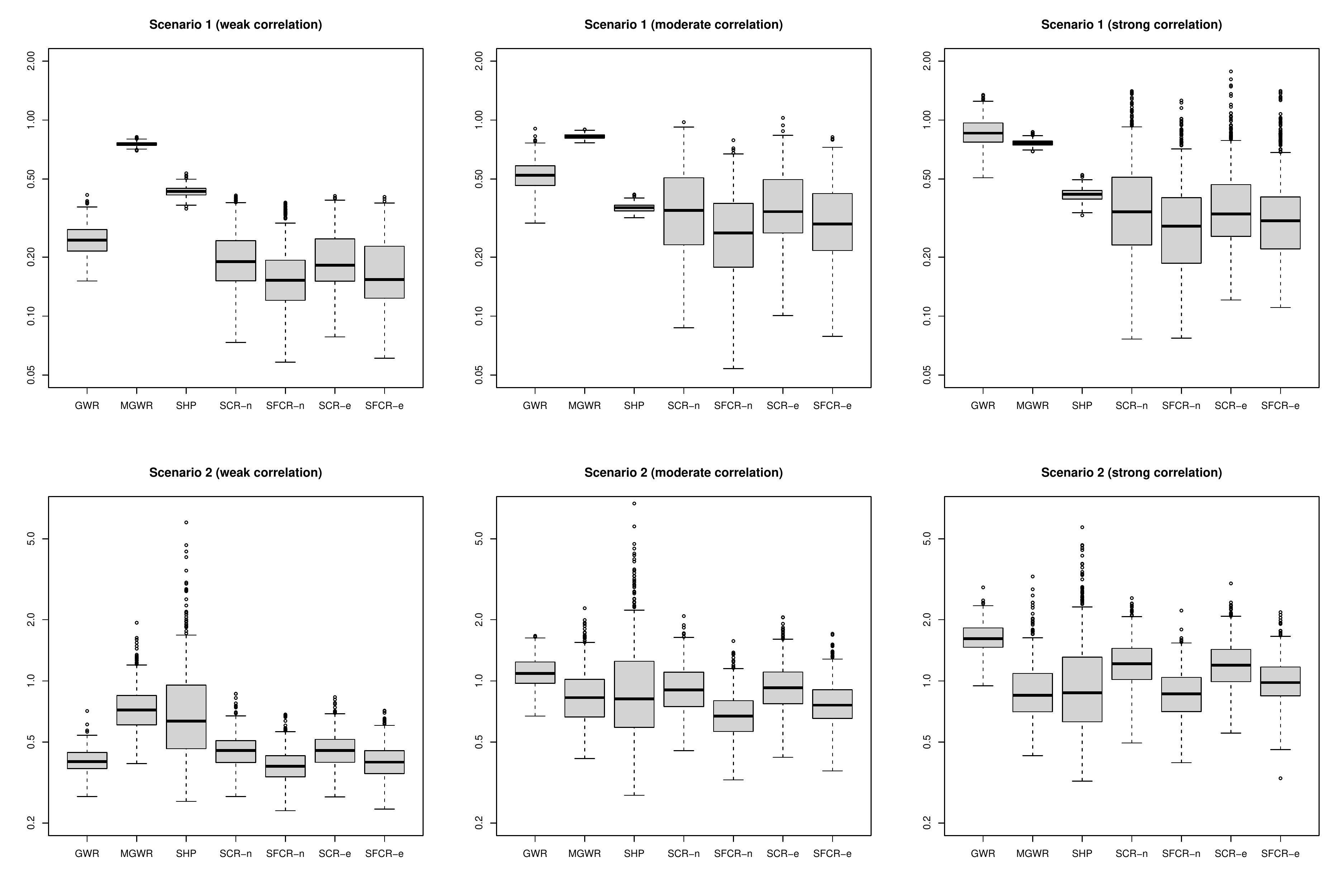}
\caption{Boxplot of MSE for GWR, SHP, SCR and SFCR based on 1000 simulated datasets. 
\label{fig:sim-box}
}
\end{figure}

We next report the estimation performance based on 1000 simulated datasets under weak ($\eta=0.2$), moderate ($\eta=0.6$) and strong ($\eta=1$) spatial correlation in covariates. 
The boxplots of MSE are given in Figure \ref{fig:sim-box}.
In scenario 1, SCR works better than GWR regardless of the strength of spatial correlation due to the underlying clustered structures of regression coefficients. 
Although SHP takes account of clustered structures, the performance is not necessarily preferable to the other methods, possibly because SHP implicitly assumes that the error variance is spatially homogeneous. 
It is also observed that the performance of GWR under moderate or strong spatial correlation in covariates is not satisfactory.
In scenario 2, the proposed methods (SCR and SFCR) and GWR are quite comparable when the spatial correlation in covariates is weak, whereas the proposed methods tend to perform better than GWR as the spatial correlation increases. 
Although the performance of SHP is not preferable under a weak spatial correlation, the relative performance gets improved as the spatial correlation increases. 
Comparing SCR (hard clustering version) and SFCR (fuzzy clustering version), SFCR can provide slightly better estimates than SCR since the true spatial patterns are smooth.

\begin{figure}[!htb]
\centering
\includegraphics[width=14cm,clip]{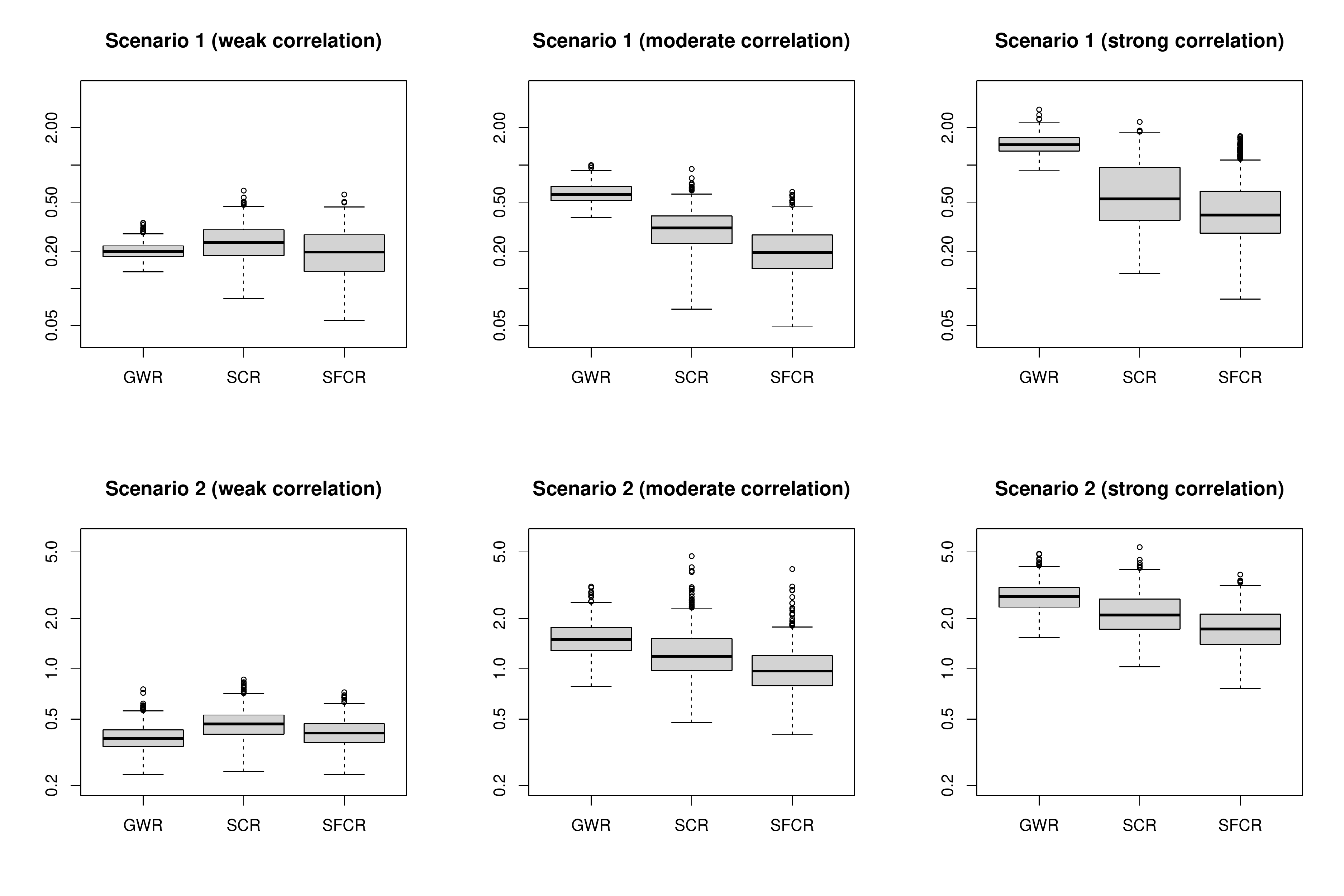}
\caption{Boxplot of MSE for GWR, SCR and SFCR in terms of spatial interpolation based on 200 simulated datasets. 
\label{fig:sim-pred-box}
}
\end{figure}

Finally, we show the results of spatial interpolation (estimation in locations without samples).
The boxplots of the MSE values based on 200 replications are shown in Figure \ref{fig:sim-pred-box}.
It is confirmed that the proposed methods provide better interpolation than GWR, especially when the correlation is moderate or strong in both scenarios. 
It should be worth noting that the proposed methods perform better than GWR even when the underlying spatial distribution of the regression coefficient is smooth.

In summary, the proposed method can produce spatially varying estimates that are more precise or as precise as those of the existing methods, while the computation time is comparable with GWR and is much shorter than SHP. 
Hence, the proposed method would be a preferable alternative for flexible spatial regression under a large spatial dataset.

\begin{table}[!htb]
\caption{Computation time (seconds) of the four methods in one simulation. 
\label{tab:CPT}
}
\begin{center}
\begin{tabular}{cccccccc}
\hline
& & GWR & SHP & SCR & SFCR \\
\hline
scenario 1 & & 4.1 & 93.0 & 3.5 & -- \\
scenario 2 & & 4.3 & 87.0 & 1.4 & 2.2 \\ 
\hline
\end{tabular}
\end{center}
\end{table}

\subsection{Computation time under large spatial data}
\begin{figure}[!htb]
\centering
\includegraphics[width=10cm,clip]{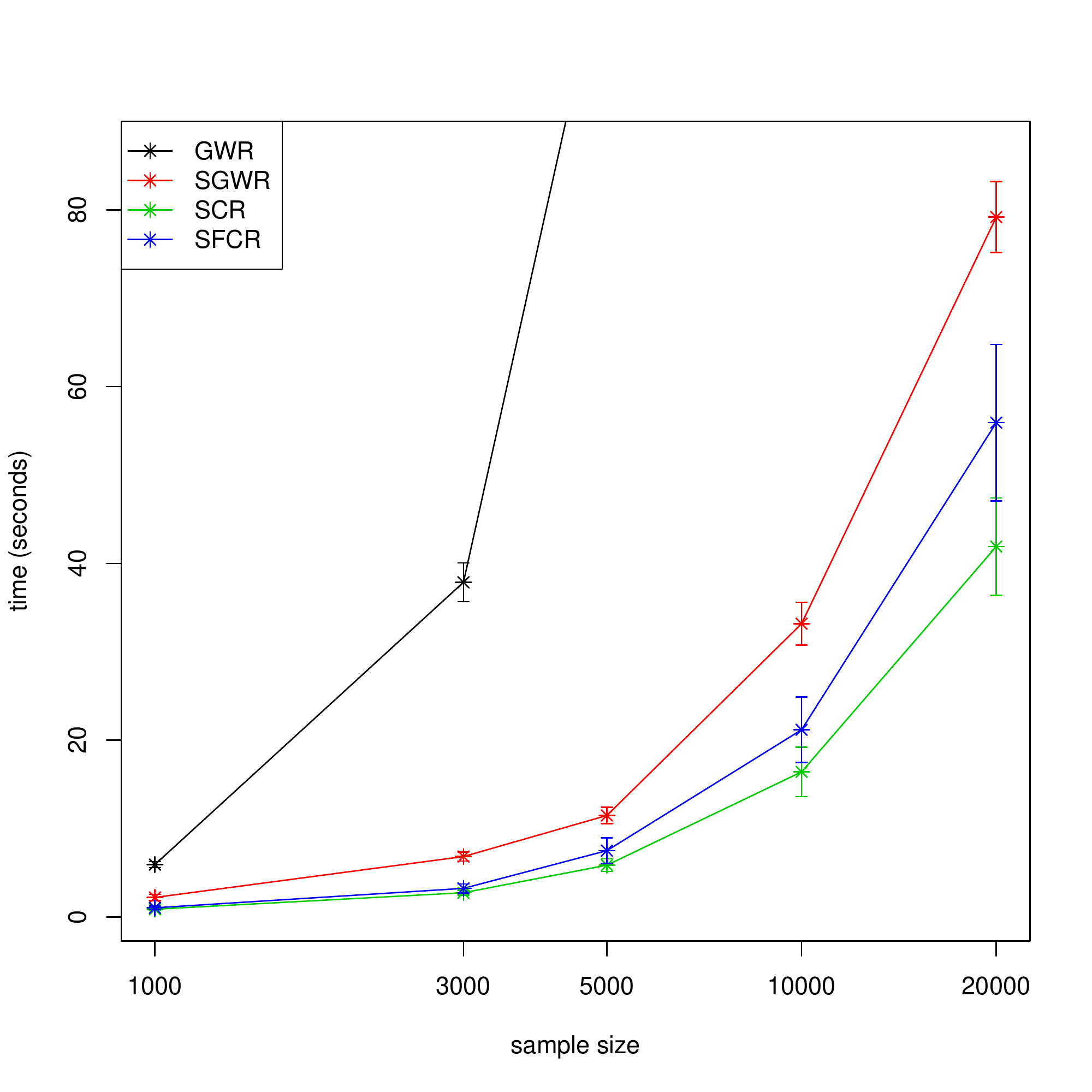}
\caption{Computation time (seconds) of GWR, SGWR, SCR and SFCR under large samples, which are averaged over 20 replications. 
The vertical lines represent double standard deviations of the replicated computation times.   
\label{fig:CPT}
}
\end{figure}

Finally, we evaluated the scalability of the proposed method under large spatial datasets.
As benchmark methods, we adopted the standard GWR and the recently proposed scalable version of GWR \citep{murakami2020scalable}, denoted by SGWR. 
In this study, we set $\eta=0$ (no spatial correlations in covariates), scenario 1 for the regression coefficients, and considered five cases of the sample size, namely, $n\in \{1000, 3000, 5000, 10000, 20000\}$. 
For each $n$, we generated 20 datasets and applied GWR, SGWR, SCR, and SFCR, where the tuning parameter was selected in the same way as in Section \ref{sec:sim-method}.
The averaged value of computation times over 20 replications and error bars representing double standard deviations are shown in Figure \ref{fig:CPT}. 
The results reveal that the computation time of GWR rapidly increases as $n$ increases.
Although the computation time of the scalable version (SGWR) is relatively shorter than the original GWR, especially under large $n$ situations, the proposed SCR and SFCR provide consistently shorter computation time than SGWR.

\section{Application to crime risk modeling}\label{sec:app}
Here we apply the proposed methods to a dataset of the number of police-recorded crime in the Tokyo metropolitan area, provided by the University of Tsukuba and publicly available online (``GIS database of the number of police-recorded crime at O-aza, chome in Tokyo, 2009-2017'', available at \url{https://commons.sk.tsukuba.ac.jp/data_en}). 
In this study, we focus on the number of violent crimes in $n=2,855$ local towns in the Tokyo metropolitan area in 2015.
For auxiliary information about each town, we adopted area (km$^2$), entire population density (PD), day-time population density (DPD), the density of foreign people (FD), percentage of single-person households (SH), and average year of living (AYL).
Let $y_i$ be the observed count of violent crimes, $s_i$ be a two-dimensional vector of longitude and litigate of the center, $a_i$ be area (km$^2$) and $x_i$ be the vector of standardized auxiliary information in the $i$th local town.
For estimating the structure of the number of crimes explained by the covariates, we employed the following spatially clustered negative binomial model:
\begin{equation}\label{NB-crime}
y_i\sim {\rm NB}(a_i\exp(x_i^t\beta_{g_i}), \nu_{g_i}), \ \ \ \ i=1,\ldots,n,
\end{equation}
where $\beta_{g_i}$ is a vector of unknown regression coefficients, $\nu_{g_i}$ is an overdispersion parameter, and ${\rm NB}(\mu, r)$ is the negative binomial distribution with mean $\mu$ and dispersion $r$. 
Under the model (\ref{NB-crime}), the expectation of $y_i/a_i$ is $\exp(x_i^t\beta_{g_i})$, so the regression term can be interpreted as the crime risk per unit km$^2$.

We first apply the proposed SCR and SFCR. 
We obtained the spatial contingency matrix based on the geographical information by choosing the five nearest locations for each location.  
Then, we set $\phi=1$ and selected the number of groups $G$ from $G\in\{1,\ldots,15\}$ based on the BIC-type criterion (\ref{IC}), and we obtained $G=7$ as the optimal choice.
With the selected $G$, we apply SCR and SFCR with $\delta=1$.
To check the sensitivity of $\delta$ in SFCR, we tried three other choices of $\delta$ ($\delta=0.5, 2$ and $5$), but the estimation results did not change very much. 
For comparisons, we applied the geographically weighted negative binomial regression \citep{da2014geographically}, denoted by GWNB. 
The Gaussian kernel is used for the weighting function, and the bandwidth is selected via cross-validation.

\begin{figure}[!htb]
\centering
\includegraphics[width=15cm,clip]{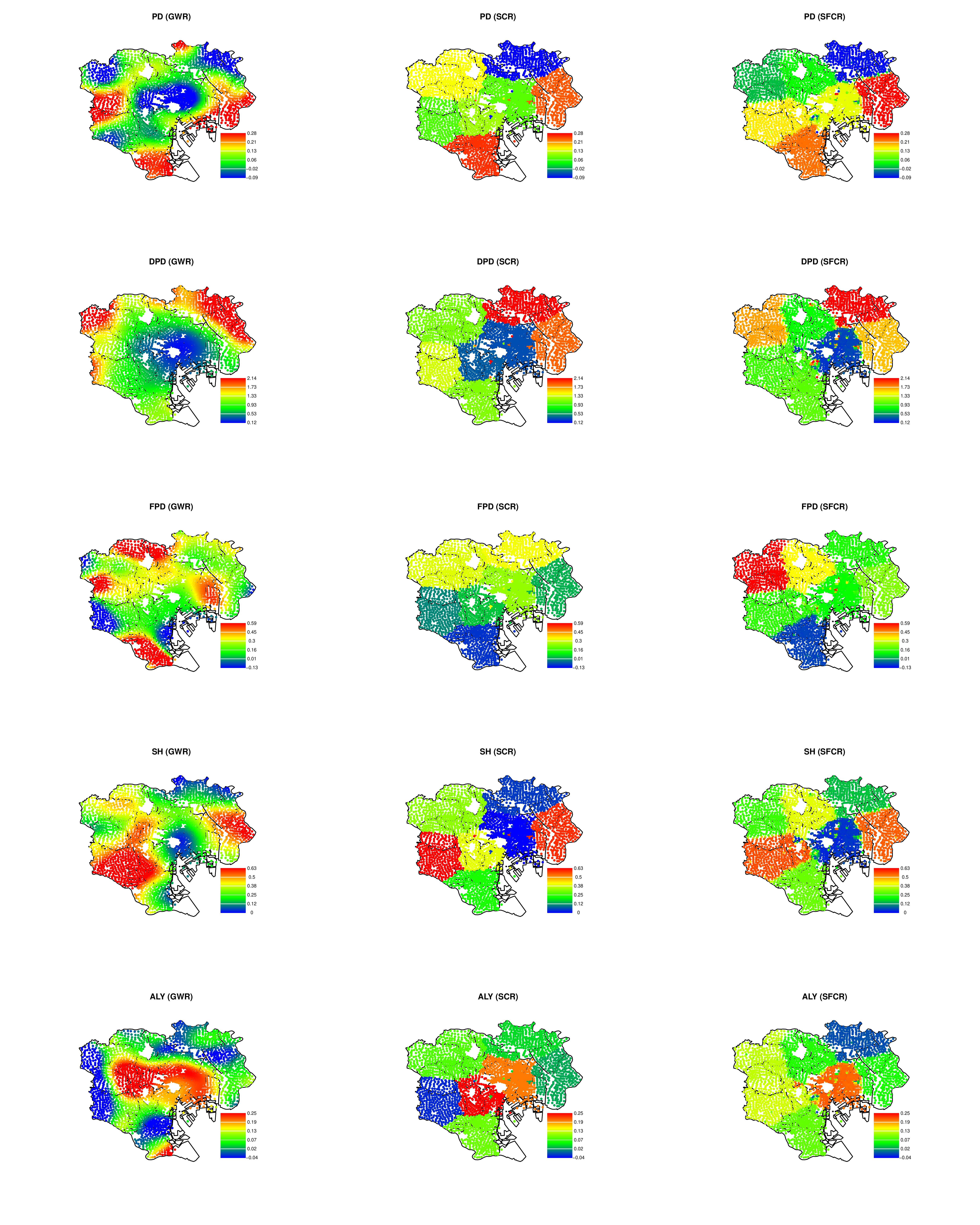}
\caption{Estimates of spatially varying coefficients for the five covariates, PD, DPD, FD, SH, and AYL based on geographically weighted negative binomial regression (GWNB), spatially clustered regression (SCR), and spatially fuzzy clustered regression (SFCR).
\label{fig:crime}
}
\end{figure}

In Figure \ref{fig:crime}, we reported the estimated spatially varying regression coefficients for all the covariates.
It is observed that GWNB produces estimates that change drastically over the space, and the change tends to be more drastic around the edge of the space.
In particular, the estimated coefficients of PD or ALY are not very smooth; thereby, the interpretation of the results is not straightforward. 
On the other hand, the proposed SCR method provides reasonable spatial clustering results and estimates of group-wise regression coefficients, and the results are highly interpretable compared with those of GWNB, while the overall spatial trend obtained from both methods is relatively similar. 
Comparing SCR and SFCR, SFCR tends to provide slightly more smoothed estimates than SCR, especially around the boundaries between clusters.
Focused on SCR and SFCR, their coefficients on PD take large values in the south and east areas while those on DPD in the northeast areas. These areas are residential areas. These results suggest high crime risk in densely populated districts (e.g., shopping districts) in these residential areas. On the other hand, the coefficients on ALY have large positive values in the central area. It is known that people tend to commit crimes in an area where they lived for a long time because they are familiar with that area \citep{bernasco2010effects}. Based on the coefficients on ALY, such a tendency is strong in the central area.

We next investigate the performance of the models in terms of prediction. 
To this end, we first randomly eliminated $m=200$ locations, which are kept as ``test data".
Using the remained training data, we estimated the regression coefficients in the omitted locations based on SCR, SFCR, and GWNB, where the same values of the number of groups or bandwidth parameters were adopted.
Then, we predicted $y_i$ using the information of $x_i$ and $a_i$, and the prediction accuracy was assessed via the following two measures: 
$$
{\rm MAPE}=\frac1m\sum_{j\in D}\frac{|\hat{y}_j-y_j|}{y_j+1}, \ \ \ \ \ \ \ \ 
{\rm RMSE}=\left\{\frac1m\sum_{j\in D}(\hat{y}_j-y_j)^2\right\},
^{1/2}
$$
where $\hat{y}_j$ is the predicted value, and $D$ is the index set for the test data.
The results are shown in Table \ref{tab:valid}, which shows that the proposed methods tend to produce more stable spatial prediction than GWNB.

Finally, we consider another design of weight, $w_{ij}$, in the proposed methods.
In addition to the spatial contingency matrix (denoted by $w_{1ij}$), we construct another contingency matrix (denoted by $w_{2ij}$) by choosing the five nearest neighbors in terms of distance of the covariate information $x_i$.
Then, we define $w_{ij}=(w_{1ij}+w_{2ij})/2$, which takes account of not only geographical closeness but also covariate similarities. 
The proposed SCR and SFCR methods with the covariate-dependent weight design are denoted by SCR-cd and SFCR-cd.  
The performance of SCR-cd and SFCR-cd are investigated through the spatial prediction using MAPE and RMSE, where the results are given in Table \ref{tab:valid}.
The results show that the prediction performance can be successfully improved by introducing the covariate-dependent design.

\begin{table}[!htb]
\caption{Two performance measures, MAPE and RMSE, of the five methods. 
\label{tab:valid}
}
\begin{center}
\begin{tabular}{cccccccc}
\hline
& & GWNB & SCR & SFCR & SCR-cd & SFCR-cd \\
\hline
MAPE & & 0.959 & 0.830 &  0.870 & 0.781 & 0.639 \\
RMSE & & 6.31 & 4.26 & 11.80 & 4.89 & 4.16\\ 
\hline
\end{tabular}
\end{center}
\end{table}

\section{Concluding remarks}\label{sec:disc}
This paper proposes a new spatial regression technique, called spatially clustered regression (SCR), accounting for spatial heterogeneity in model parameters by explicitly introducing grouping parameters.  
By employing a penalty function motivated by the Potts model, we formulated the penalized likelihood function easily maximized via a simple iterative algorithm. 
We also developed a fuzzy version of the method that can produce more spatially smoothed estimates and considered straightforward but essential extensions of the main idea. 
Compared with the most standard technique, GWR, we numerically confirmed that SCR performs better than or as well as GWR in terms of parameter estimation, and the computational cost of SCR is much smaller than that of GWR under large spatial data.

We finally discuss two meaningful extensions of the proposed method.
The first one is variable selection by incorporating regularization techniques into the objective function (\ref{PL}), given by 
\begin{equation}\label{lasso}
\sum_{i=1}^n \log f(y_i|x_i;\theta_{g_i})+\phi\sum_{i<j}w_{ij}I(g_i=g_j)-\lambda \sum_{g=1}^G\sum_{k=1}^p{\rm pen}(\theta_{gk}),
\end{equation}
where $\lambda$ is a tuning parameter and ${\rm pen}(\cdot)$ is a penalty function, e.g. ${\rm pen}(x)=|x|$ for Lasso regularization \citep{Tib1996}.
Under the formulation, the updating step for $\theta$ in Algorithm 1 is changed as follows: 
$$
\theta_g^{(k+1)}=\argmax_{\theta_g} \left\{\sum_{i=1}^nI(g_i^{(k)}=g)\log f(y_i|x_i;\theta_g)-\lambda \sum_{k=1}^p{\rm pen}(\theta_{gk})\right\}.
$$ 
The above objective function is the same as the penalized log-likelihood based only on the samples classified in the $g$th group; thereby, existing efficient computation algorithms could be applied to update $\theta_g$.
It should be noted that the use of the objective function (\ref{lasso}) leads to different selected variables in each group. 
Thus (\ref{lasso}) does not necessarily induce variable selection of the $p$ variables in $x_i$.
It might be more beneficial to determine the variable which is not used in all the $G$ models in practice.
To this end, we also suggest using a grouped penalty function $\sum_{k=1}^p{\rm pen}(\theta_{1k},\ldots,\theta_{Gk})$ instead of the element-wise penalty adopted in (\ref{lasso}), where ${\rm pen}(\theta_{1k},\ldots,\theta_{Gk})$ is the simultaneous penalty on $G$ regression coefficients of the $k$th variable. 
A standard choice would be grouped lasso penalty \citep{Yuan2006} given by ${\rm pen}(\theta_{1k},\ldots,\theta_{Gk})=\sqrt{\sum_{g=1}^G\theta_{gk}^2}$.
Following \cite{Zhou2007}, we can modify the information criterion (\ref{IC}) to select the tuning parameter $\lambda$, that is, we replace ${\rm dim}(\theta)$ with the degrees of freedom $\sum_{g=1}^G\sum_{k=1}^pI(\widehat{\theta}_{gk}\neq 0)$ in (\ref{IC}).
In this case, the information criterion is a function of both $G$ and $\lambda$.

The second extension is to handle semiparametric structures for the regression part.
Suppose the conditional distribution is expressed as $f(y_i|x_i;H_{g_i}, \gamma_{g_i})$, where $H_g=\{h_{g1},\ldots,h_{gp}\}$ is a collection of unknown $p$ functions and $\gamma$ is a dispersion parameter.  
For example, the linear additive model is expressed as $y_i\sim N(\sum_{k=1}^ph_{g_ik}(x_{ik}), \sigma_{g_i}^2)$, so that $E[y_i|x_i]=\sum_{k=1}^ph_{g_ik}(x_{ik})$.  
In the model, the additive effect of each covariate can be different among $G$ groups, and the model can be seen as a semiparametric version of the model discussed in Section \ref{sec:method}. 
The estimation of the model can be done via a slight modification of Algorithms 1 and 2. The updating step for $\theta_g$ is replaced with one for $H_g$, given by 
$$
H^{(k+1)}_g=\argmax_{H_g} \sum_{i=1}^nI(g_i^{(k)}=g)\log f(y_i|x_i;H_g,\gamma_g^{(k)}).
$$
The above optimization step is nothing but fitting the generalized additive models for observations assigned to the $g$th group, so that standard techniques such as sequential fitting \citep[e.g.][]{Hastie1986} can be adopted to obtain $H^{(k+1)}_g$. 
The dispersion parameter $\gamma_g$ can be updated in the same manner.
The two extensions, as mentioned above, would be helpful in practice, but the detailed theoretical and numerical investigation is left to future works.

\section*{Acknowledgements}
This work was supported by the Japan Society for the Promotion of Science (KAKENHI) Grant Numbers 18K12757 and 18H03628.

\vspace{5mm}
\bibliography{refs2}
\bibliographystyle{chicago}

\end{document}